\documentclass{ws-mpla}

\usepackage{bm}

\def\be{\begin{equation}}
\def\ee{\end{equation}}
\def\bea{\begin{eqnarray}}
\def\eea{\end{eqnarray}}
\def\lb{\label}

\begin{document}

\markboth{B. Linet and P. Teyssandier}
{Quantum phase shift and neutrino oscillations}

\catchline{}{}{}{}{}

\title{QUANTUM PHASE SHIFT AND NEUTRINO OSCILLATIONS\\
IN A STATIONARY, WEAK GRAVITATIONAL FIELD
}

\author{\footnotesize B. LINET}

\address{Laboratoire de Math\'ematiques et Physique Th\'eorique, 
CNRS/UMR 6083,\\ F\'ed\'eration Denis Poisson, Universit\'e Fran\c{c}ois Rabelais, F-37200 Tours,
France\\
Bernard.Linet@lmpt.univ-tours.fr}

\author{P. TEYSSANDIER}

\address{D\'epartement Syst\`emes de R\'ef\'erence Temps Espace, 
CNRS/UMR 8630, UPMC,\\ Observatoire de Paris, 61 avenue de l'Observatoire,
F-75014 Paris, France\\
Pierre.Teyssandier@obspm.fr
}

\maketitle

\pub{}{}

\begin{abstract}
A new method based on Synge's world function is developed for determining within the WKB approximation the gravitationally 
induced quantum phase shift of a particle propagating in a stationary spacetime. This method avoids any calculation of geodesics. A detailed treatment is given for relativistic particles within the weak field, linear approximation of any metric theory. The method is applied to the calculation of the oscillation terms governing the interference of neutrinos considered as the superposition of two eigenstates having different masses. It is shown that the neutrino oscillations are not sensitive to the gravitomagnetic components of the metric as long as the spin contributions can be ignored. Explicit calculations are performed when the source of the field is a spherical, homogeneous body. A comparison is made with previous results obtained in Schwarzschild spacetime.

\keywords{Gravitation field; quantum phase; neutrinos.}
\end{abstract}

\ccode{PACS Nos.: 04.25.-g; 95.30.Sf; 14.60.Pq.
}

\section{Introduction}

A lot of papers has been devoted to the effects of gravitation on the neutrino oscillations 
(see, e.g., Refs.~\refcite{Ahluwalia1}-\refcite{Capozziello2}). 
As far as we know, all the proposed derivations of the neutrino oscillation formula 
within the WKB approximation require some integrations along null or timelike geodesics. For this reason, even in the simple case of the Schwarzschild metric treated in the linear approximation, the calculations become heavy for non-radial propagations and the results 
remain limited to the case where the neutrinos are created outside the matter generating the gravitational field. 
The purpose of the present paper is to provide a new method which avoids any determination of geodesics. Based on Synge's world function as defined in Ref.~\refcite{Synge}, this method has shown its usefulness in the study of light rays (see Refs.~\refcite{Linet1}-\refcite{Teyssandier}) and can be applied to neutrinos and more generally to all kinds of particles, as long as the spin 
contributions may be ignored.\footnote{Note that the gravitational contributions to the spin precession 
vanish in the Schwarzschild spacetime (see, e.g., Refs.~\refcite{Piriz} and \refcite{Cardall}).}

We begin by giving a general relation between the quantum phase of a particle freely propagating 
in a stationary spacetime and Synge's world function. Then, restricting our attention to the case of relativistic particles, we obtain the formulas yielding the travel time and the quantum accumulated phase within the framework of the weak field, linear approximation 
of any metric theory of gravity. It must be noted that our method is not suitable for the cases where multiple-path propagation effects may occur since the definition of Synge's world function requires the existence of a single geodesic path between two given points-events.

We apply our method to the determination of the oscillations of neutrinos considered as superpositions of two eigenstates 
with different masses. We assume that the different eigenstates of the 
neutrinos are described by wave packets and that only the components of these wave packets having the same energy 
contribute to the interference (see, e.g., Refs.~\refcite{Stodolsky1}, \refcite{Lipkin} and \refcite{Crocker}). For practical calculations we consider that the metric perturbations are given by the terms 
of order $G/c^2$ and $G/c^3$ within the standard post-Newtonian formalism, $G$ being the Newtonian 
gravitational constant. 

We obtain explicit expressions for a gravitational field generated by a spherical, homogeneous body. Our results hold 
even in the case of non-radial propagations and for sources located inside the central body. Finally, we compare our formula giving the accumulated quantum phase with the results found in Ref.~\refcite{Crocker} for the exterior Schwarzschild metric. 

We suppose that spacetime is covered by a global quasi Minkowskian coordinate 
system $(x^{\mu})$. We put $x^0 = ct$, $t$ being a timelike coordinate and we use the notation 
${\bf x} =(x^i)$. The signature adopted for the metric tensor $g_{\mu\nu}$ is $(+---)$.

\section{Phase shift in a weak stationary gravitational field}

\subsection{Quantum phase shift and world function}  

Let us consider a free particle with a mass $m$ propagating from a point $x_{\!\scriptscriptstyle A}$ to a 
point $x_{\!\scriptscriptstyle B}$. The quantum phase of this particle accumulated between $x_{\!\scriptscriptstyle A}$ and 
$x_{\!\scriptscriptstyle B}$ is given by
\be \lb{2}
\Phi_{\!\scriptscriptstyle AB}=\frac{mc}{\hbar}s_{\!\scriptscriptstyle AB} \, ,   
\ee
where $s_{\!\scriptscriptstyle AB}$ is the length of the timelike geodesic $\Gamma_{\!\scriptscriptstyle AB}$ joining $x_{\!\scriptscriptstyle A}$ 
and $x_{\!\scriptscriptstyle B}$ (see, e.g., Ref.~\refcite{Stodolsky2}). Let us introduce Synge's world function $\Omega (x_{\!\scriptscriptstyle A}, x_{\!\scriptscriptstyle B}) $, defined as
\be \lb{t2b}
\Omega (x_{\!\scriptscriptstyle A}, x_{\!\scriptscriptstyle B}) = \frac{1}{2} \int_{0}^{1} g_{\mu\nu}(x^{\alpha}(\lambda))\frac{dx^{\mu}}{d\lambda}
\frac{dx^{\nu}}{d\lambda} d\lambda,
\ee
where the integral is taken over $\Gamma_{\!\scriptscriptstyle AB}$ and $\lambda$ is the unique affine parameter along $\Gamma_{\!\scriptscriptstyle AB}$ 
which satisfies the boundary conditions $\lambda_{\!\scriptscriptstyle A} = 0$ and $\lambda_{\!\scriptscriptstyle B} = 1$. Since 
$s_{\!\scriptscriptstyle AB}^2 = 2 \, \Omega (x_{\!\scriptscriptstyle A}, x_{\!\scriptscriptstyle B})$ (cf. Refs.~\refcite{Synge} and \refcite{Linet2}), $\Phi_{\!\scriptscriptstyle AB}$ and $\Omega (x_{\!\scriptscriptstyle A}, x_{\!\scriptscriptstyle B})$ are linked by the relation
\be \lb{2a}
\Phi_{\!\scriptscriptstyle AB}=\frac{mc}{\hbar}\sqrt{2\Omega (x_{\!\scriptscriptstyle A}, x_{\!\scriptscriptstyle B})} \, .
\ee

As a consequence, the knowledge of the world function corresponding to a given metric 
enables to determine the accumulated phase $\Phi_{\!\scriptscriptstyle AB}$. In addition, the world function enables to 
perform the calculation of the energy-momentum vector $p_{\alpha} = m c u_{\alpha}$ of the particle 
at $x_{\!\scriptscriptstyle A}$ and $x_{\!\scriptscriptstyle B}$, $u_{\alpha}$ denoting the covariant components of the unit 4-velocity vector $dx^{\alpha}/ds$ along 
$\Gamma_{\!\scriptscriptstyle AB}$. Indeed, it may be inferred from the definition of the world function that the values of $u_{\alpha}$ at $x_{\!\scriptscriptstyle A}$ and $x_{\!\scriptscriptstyle B}$ are given by (see Refs.~\refcite{Synge} and \refcite{Linet2}) 
\be \lb{t2c}
(u_{\alpha})_{\!\scriptscriptstyle A} = - \frac{1}{\sqrt{2\Omega (x_{\!\scriptscriptstyle A}, x_{\!\scriptscriptstyle B})}} \frac{\partial \Omega}{\partial x^{\alpha}_{\!\scriptscriptstyle A}} \, , 
\quad (u_{\alpha})_{\!\scriptscriptstyle B} = \frac{1}{\sqrt{2\Omega (x_{\!\scriptscriptstyle A}, x_{\!\scriptscriptstyle B})}} \frac{\partial \Omega}{\partial x^{\alpha}_{\!\scriptscriptstyle B}}\, .
\ee

\subsection{Phase shift in a stationary spacetime}  

Henceforth, we restrict our attention to stationary spacetimes. 
We choose the coordinate system $x^{\mu}$ so that the metric components do not depend on $x^0$. The world function then has the
form $\Omega (x_{\!\scriptscriptstyle B}^{0}-x_{\!\scriptscriptstyle A}^{0},{\bf x}_{\!\scriptscriptstyle A},{\bf x}_{\!\scriptscriptstyle B})$ and the component $u_0$ is a constant of the 
motion, i.e. 
\be \lb{t2d}
u_0 = E/mc^2 \, , \quad E = \mbox{const}.
\ee 
Let ${\bf x}$ be a point through which the particle is passing. The energy of the particle as locally measured by a stationary 
observer staying at ${\bf x}$ is the quantity given by 
\be \lb{Elo}
E_{loc}({\bf x}) = \frac{E}{\sqrt{g_{00}({\bf x})}}.
\ee

It is easily deduced from Eqs. (\ref{t2c}) written for $\alpha=0$ and from Eq. (\ref{t2d}) that $x_{\!\scriptscriptstyle B}^{0}-x_{\!\scriptscriptstyle A}^{0}$, ${\bf x}_{\!\scriptscriptstyle A}$, ${\bf x}_{\!\scriptscriptstyle B}$ and $E$ are 
linked by the relation 
\be \lb{n1}
\sqrt{2\Omega (x_{\!\scriptscriptstyle B}^{0}-x_{\!\scriptscriptstyle A}^{0},{\bf x}_{\!\scriptscriptstyle A},{\bf x}_{\!\scriptscriptstyle B})}=\frac{mc^2}{E}\dot{ \Omega}(x_{\!\scriptscriptstyle B}^{0}-x_{\!\scriptscriptstyle A}^{0},{\bf x}_{\!\scriptscriptstyle A},{\bf x}_{\!\scriptscriptstyle B}),
\ee
where the following notation is used
\be  \lb{dO} 
\dot{ \Omega}(x_{\!\scriptscriptstyle B}^{0}-x_{\!\scriptscriptstyle A}^{0},{\bf x}_{\!\scriptscriptstyle A},{\bf x}_{\!\scriptscriptstyle B})\equiv\frac{\partial \Omega (x_{\!\scriptscriptstyle B}^{0}-x_{\!\scriptscriptstyle A}^{0},{\bf x}_{\!\scriptscriptstyle A},{\bf x}_{\!\scriptscriptstyle B})}{\partial (x_{\!\scriptscriptstyle B}^{0}-x_{\!\scriptscriptstyle A}^{0})}.
\ee

Solving Eq. (\ref{n1}) for $x_{\!\scriptscriptstyle B}^{0}-x_{\!\scriptscriptstyle A}^{0}$ gives the (coordinate) travel time of the particle between 
its point of emission and its point of reception as a function of ${\bf x}_{\!\scriptscriptstyle A}$, ${\bf x}_{\!\scriptscriptstyle B}$ and $E$ for a given mass $m$. In what follows we shall denote 
this function by ${\cal T}({\bf x}_{\!\scriptscriptstyle A}, {\bf x}_{\!\scriptscriptstyle B}, E; m)$, so that $x_{\!\scriptscriptstyle B}^{0}-x_{\!\scriptscriptstyle A}^{0}$ is given by
\be \lb{dT}
x_{\!\scriptscriptstyle B}^{0}-x_{\!\scriptscriptstyle A}^{0}=c{\cal T}({\bf x}_{\!\scriptscriptstyle A}, {\bf x}_{\!\scriptscriptstyle B}, E; m).
\ee

Substituting for $x_{\!\scriptscriptstyle B}^{0}-x_{\!\scriptscriptstyle A}^{0}$ from Eq. (\ref{dT}) into the right-hand side (R.H.S.) of Eq. (\ref{n1}) yields $\sqrt{2\Omega}$ as a function of ${\bf x}_{\!\scriptscriptstyle A}$, ${\bf x}_{\!\scriptscriptstyle B}$, $E$ and $m$. Consequently, Eq. (\ref{2a}) reads
\be \lb{n2}
\Phi_{\!\scriptscriptstyle AB}=\frac{m^2c^3}{\hbar E}\dot{ \Omega}(c{\cal T}({\bf x}_{\!\scriptscriptstyle A}, {\bf x}_{\!\scriptscriptstyle B}, E; m), {\bf x}_{\!\scriptscriptstyle A}, {\bf x}_{\!\scriptscriptstyle B}).
\ee

Noting that Eq. (\ref{2}) is equivalent to (see, e.g., Ref.~\refcite{Stodolsky2})
\be \lb{Sto}
\Phi_{\!\scriptscriptstyle AB}=\frac{1}{\hbar}\int_{\Gamma_{\!\scriptscriptstyle AB}}p_{\mu}dx^{\mu}
\ee
and that $p_0$ is a constant of the motion given by $p_0=E/c$, the accumulated phase may be written in the form
\be  \lb{Sto2}
\Phi_{\!\scriptscriptstyle AB}=\frac{E}{\hbar c}(x_{\!\scriptscriptstyle B}^{0}-x_{\!\scriptscriptstyle A}^{0}) + \Psi_{\!\scriptscriptstyle AB},
\ee
where $\Psi_{\!\scriptscriptstyle AB}=1/\hbar\int_{{\!\scriptscriptstyle A}}^{{\!\scriptscriptstyle B}}p_idx^i$ is a function of ${\bf x}_{\!\scriptscriptstyle A}$, ${\bf x}_{\!\scriptscriptstyle B}$, $E$ and $m$. Indeed, inserting Eqs. (\ref{dT}) 
and (\ref{n2}) into Eq. (\ref{Sto2}), it is easily seen that
\be \lb{Psi1}
\Psi_{\!\scriptscriptstyle AB}=\frac{E}{\hbar c}\left[\frac{m^2c^4}{E^2}\dot{ \Omega}(c{\cal T}({\bf x}_{\!\scriptscriptstyle A}, {\bf x}_{\!\scriptscriptstyle B}, E; m), {\bf x}_{\!\scriptscriptstyle A}, {\bf x}_{\!\scriptscriptstyle B}) - c{\cal T}({\bf x}_{\!\scriptscriptstyle A}, {\bf x}_{\!\scriptscriptstyle B}, E; m)
\right].
\ee

We shall see below that $\Psi_{\!\scriptscriptstyle AB}$ is the relevant quantity for determining the stationary oscillations of neutrinos. As a consequence, Eq. (\ref{Psi1}) constitutes the basic formula of the present work and we shall henceforth concentrate our attention on the determination of its right-hand side.

\subsection{Weak field, linear approximation} 

Let us now specialize in the weak field, linear approximation. We assume that there exists a coordinate system 
in which the metric may be written in the form
\be\lb{1}
g_{\mu \nu} = \eta_{\mu \nu} + h_{\mu \nu}+O(G^2),
\ee
where $\eta_{\mu \nu} = $ diag $(1, -1, -1 ,-1)$ and the quantities $h_{\mu \nu}$ are 
considered as perturbation terms of the first order in $G$. It is shown in Ref.~\refcite{Linet1} that the 
world function is then given by 
\be\lb{4}
\Omega (x_{\!\scriptscriptstyle A}, x_{\!\scriptscriptstyle B}) = \frac{1}{2}\left[\eta_{\mu\nu}+\int_{0}^{1}h_{\mu \nu}(x_{(0)}(\lambda )) d\lambda\right]
(x_{\!\scriptscriptstyle B}^{\mu}-x_{\!\scriptscriptstyle A}^{\mu})(x_{\!\scriptscriptstyle B}^{\nu}-x_{\!\scriptscriptstyle A}^{\nu}) + O(G^2),
\ee
the integration being performed along the straight line of parametrized equations 
$x_{(0)}^{\alpha}(\lambda )=(x_{\!\scriptscriptstyle B}^{\alpha}-x_{\!\scriptscriptstyle A}^{\alpha})\lambda +x_{\!\scriptscriptstyle A}^{\alpha}$. Since spacetime is assumed to be stationary, 
Eq. (\ref{4}) may be written as 
\bea \lb{8}
\Omega (x_{\!\scriptscriptstyle A},x_{\!\scriptscriptstyle B}) & = & \frac{1}{2} (x_{\!\scriptscriptstyle B}^{0}-x_{\!\scriptscriptstyle A}^{0})^2 \left[ 1 + 2{\mathcal A}({\bf x}_{\!\scriptscriptstyle A},{\bf x}_{\!\scriptscriptstyle B})\right]
+(x_{\!\scriptscriptstyle B}^{0}-x_{\!\scriptscriptstyle A}^{0}) R_{\!\scriptscriptstyle AB} {\mathcal B}({\bf x}_{\!\scriptscriptstyle A},{\bf x}_{\!\scriptscriptstyle B})    \nonumber \\ 
&  &- \frac{1}{2} R_{\!\scriptscriptstyle AB}^{2}\left[1 - 2 {\mathcal C}({\bf x}_{\!\scriptscriptstyle A},{\bf x}_{\!\scriptscriptstyle B}) \right] + O(G^2),
\eea
where $R_{\!\scriptscriptstyle AB}=\mid \! {\bf x}_{\!\scriptscriptstyle B}-{\bf x}_{\!\scriptscriptstyle A}\! \mid$ and
\bea
& & {\mathcal A}({\bf x}_{\!\scriptscriptstyle A},{\bf x}_{\!\scriptscriptstyle B}) = \frac{1}{2} \int_{0}^{1}h_{00}({\bf x}_{(0)}(\lambda ))d\lambda ,
\lb{A} \\
& & {\mathcal B}({\bf x}_{\!\scriptscriptstyle A},{\bf x}_{\!\scriptscriptstyle B}) = \frac{(x_{\!\scriptscriptstyle B}^{i} - x_{\!\scriptscriptstyle A}^{i})}{R_{\!\scriptscriptstyle AB}} \int_{0}^{1}h_{0i}({\bf x}_{(0)}
(\lambda ))d\lambda, \lb{B}\\
& & {\mathcal C}({\bf x}_{\!\scriptscriptstyle A},{\bf x}_{\!\scriptscriptstyle B}) = \frac{(x_{\!\scriptscriptstyle B}^{i} - x_{\!\scriptscriptstyle A}^{i})(x_{\!\scriptscriptstyle B}^{j} - x_{\!\scriptscriptstyle A}^{j})}{2 R_{\!\scriptscriptstyle AB}^{2}}
\int_{0}^{1}h_{ij}({\bf x}_{(0)}(\lambda ))d\lambda, \lb{9}
\eea
with 
\be \lb{t9a}
{\bf x}_{(0)}(\lambda )=({\bf x}_{\!\scriptscriptstyle B}-{\bf x}_{\!\scriptscriptstyle A})\lambda +{\bf x}_{\!\scriptscriptstyle A}.
\ee

For the sake of brevity, we shall henceforth omit the symbol $O(G^2)$ standing for the 
post-post-Minkowskian terms. 

Differentiating the R.H.S. of Eq. (\ref{8}) with respect to 
$x_{\!\scriptscriptstyle B}^{0}-x_{\!\scriptscriptstyle A}^{0}$, and then substituting the obtained expression into Eq. (\ref{dO}) yield 
\be \lb{n3}
\dot{ \Omega}(x_{\!\scriptscriptstyle B}^{0}-x_{\!\scriptscriptstyle A}^{0},{\bf x}_{\!\scriptscriptstyle A},{\bf x}_{\!\scriptscriptstyle B})=(x_{\!\scriptscriptstyle B}^{0}-x_{\!\scriptscriptstyle A}^{0})
\left[ 1 + 2 {\mathcal A}({\bf x}_{\!\scriptscriptstyle A},{\bf x}_{\!\scriptscriptstyle B})\right] + R_{\!\scriptscriptstyle AB} {\mathcal B}({\bf x}_{\!\scriptscriptstyle A},{\bf x}_{\!\scriptscriptstyle B}).
\ee

Squaring each side of Eq. (\ref{n1}), and then using Eqs. (\ref{8}) and (\ref{n3}) give an equation of the second degree in $x_{\!\scriptscriptstyle B}^{0}-x_{\!\scriptscriptstyle A}^{0}$. Solving 
this equation leads to\footnote{Let $m=0$ in Eq. (\ref{11}). It is easily seen that the expression of the function ${\cal T}({\bf x}_{\!\scriptscriptstyle A}, {\bf x}_{\!\scriptscriptstyle B}, E; 0)$ is independent of $E$ and coincides with the expression of the time transfer function giving the travel time of a photon between points ${\bf x}_{\!\scriptscriptstyle A}$ and ${\bf x}_{\!\scriptscriptstyle B}$ (see, e.g., Refs.~\refcite{Linet1} and \refcite{Teyssandier}).} 
\bea \lb{11}
& & c{\cal T}({\bf x}_{\!\scriptscriptstyle A}, {\bf x}_{\!\scriptscriptstyle B}, E; m)=R_{\!\scriptscriptstyle AB} \Bigg\lbrace \frac{E}
{\sqrt{E^2 - m^2c^4}}
\Bigg\lbrack 1- \left(1 - \frac{ m^2c^4}{E^2 - m^2 c^4} \right)
{\mathcal A}({\bf x}_{\!\scriptscriptstyle A},{\bf x}_{\!\scriptscriptstyle B})\nonumber \\
& &\qquad\qquad\qquad\qquad\;\, - \,{\mathcal C}({\bf x}_{\!\scriptscriptstyle A},{\bf x}_{\!\scriptscriptstyle B}) \Bigg\rbrack - {\mathcal B}({\bf x}_{\!\scriptscriptstyle A},{\bf x}_{\!\scriptscriptstyle B}) \Bigg\rbrace.
\eea

Finally, inserting Eqs. (\ref{n3}) and (\ref{11}) into Eq. (\ref{Psi1}) gives for $\Psi_{\!\scriptscriptstyle AB}$ 
\bea \lb{N5}
& &\Psi_{\!\scriptscriptstyle AB}=-R_{\!\scriptscriptstyle AB}\frac{\sqrt{E^2-m^2c^4}}{\hbar c}\Bigg\lbrack1-\frac{E^2}{E^2-m^2c^4}{\mathcal A}({\bf x}_{\!\scriptscriptstyle A}, {\bf x}_{\!\scriptscriptstyle B}) - {\mathcal C}({\bf x}_{\!\scriptscriptstyle A}, {\bf x}_{\!\scriptscriptstyle B})\nonumber \\
& &\qquad\quad-\frac{E}{\sqrt{E^2 - m^2c^4}}{\mathcal B}({\bf x}_{\!\scriptscriptstyle A}, {\bf x}_{\!\scriptscriptstyle B})\Bigg\rbrack. 
\eea

It may be seen that the expression of $c{\cal T}({\bf x}_{\!\scriptscriptstyle A}, {\bf x}_{\!\scriptscriptstyle B}, E; m)$ given by Eq. (\ref{11}) is valid only if the term $1-m^2c^4/E^2$ is significantly far from 0. So we shall henceforth assume that the condition
\be \lb{rel}
mc^2 \ll E
\ee 
is satisfied. Since it may be shown that the constant $E$ is linked to the  velocity $v_{{\bf x}}$ of the particle as measured by a stationary observer at ${\bf x}$ by the relation (cf., e.g., Ref.~\refcite{Nandi})
\be \lb{Ev}
E=mc^2\sqrt{\frac{g_{00}({\bf x})}{1-v_{{\bf x}}^2/c^2}},
\ee
we can consider that our formulas hold for relativistic particles.

In a general stationary spacetime, the infinitesimal spatial distance $d\sigma$ between two points $(x^{\alpha})$ and  $(x^{\alpha}+dx^{\alpha})$ as measured  by a stationary observer is determined by (cf. Ref.~\refcite{Landau})
\be \lb{3ds}
d\sigma^2=-\left(g_{ij}-\frac{g_{0i}g_{0j}}{g_{00}}\right)dx^idx^j.
\ee

So an intrinsic distance $D_{\!\scriptscriptstyle AB}$ between ${\bf x}_{\!\scriptscriptstyle A}$ and ${\bf x}_{\!\scriptscriptstyle B}$ may be defined by
\be \lb{DAB}
D_{\!\scriptscriptstyle AB}=\int_{{\bf x}_{\!\scriptscriptstyle A}}^{{\bf x}_{\!\scriptscriptstyle B}} d\sigma,
\ee
where the integral is taken along the geodesic path $\Gamma_{\!\scriptscriptstyle AB}^{\ast}$ relative to the metric (\ref{3ds}) joining ${\bf x}_{\!\scriptscriptstyle A}$ and ${\bf x}_{\!\scriptscriptstyle B}$. Within the linear approximation $\Gamma_{\!\scriptscriptstyle AB}^{\ast}$ is a curve given by a position function of the form
${\bf x}(\lambda)={\bf x}_{(0)}(\lambda )+{\bm \xi}(\lambda)$, where ${\bf x}_{(0)}(\lambda)$ is defined by Eq. (\ref{t9a}) and ${\bm \xi}(\lambda)$ is a vector function of order $G$ satisfying the boundary conditions
${\bm \xi}(0)=0, {\bm \xi}(1)=0$. Taking these boundary conditions into account, it is easily seen that Eqs. (\ref{3ds}) and (\ref{DAB}) yield
\be \lb{DAB2}
D_{\!\scriptscriptstyle AB}=R_{\!\scriptscriptstyle AB}[1-{\mathcal C}({\bf x}_{\!\scriptscriptstyle A}, {\bf x}_{\!\scriptscriptstyle B})]+O(G^2).
\ee

This formula generalizes the expression of the geodesic radial distance in Schwarzschild spacetime used in some discussions (see, e.g., Refs.~\refcite{Fornengo} and \refcite{Crocker}). Taking Eq. (\ref{DAB2}) into account, Eq. (\ref{N5}) reads now
\bea \lb{N5b}
& &\Psi_{\!\scriptscriptstyle AB}=-D_{\!\scriptscriptstyle AB}\frac{\sqrt{E^2 - m^2c^4}}{\hbar c}\Bigg\lbrack 1-\frac{E^2}{E^2-m^2c^4}{\mathcal A}({\bf x}_{\!\scriptscriptstyle A}, {\bf x}_{\!\scriptscriptstyle B}) \nonumber \\
& &\qquad\quad-\, \frac{E}{\sqrt{E^2 - m^2c^4}}{\mathcal B}({\bf x}_{\!\scriptscriptstyle A}, {\bf x}_{\!\scriptscriptstyle B})\Bigg\rbrack.
\eea

Owing to the shortness of the formula (\ref{N5b}), we shall systematically use $D_{\scriptscriptstyle AB}$ instead of $R_{\scriptscriptstyle AB}$ in what follows.

\section{Application to the neutrino oscillations} 

\subsection{General formulas}

Let us apply the previous results to the oscillation of neutrinos emitted at point ${\bf x}_{\!\scriptscriptstyle A}$ and detected at point ${\bf x}_{\!\scriptscriptstyle B}$. We consider that the state of a neutrino is a superposition of two eigenstates created with the same initial phase. The masses  corresponding to these eigenstates are denoted by $m_1$ and $m_2$ and 
supposed to be distinct. It results from Eq. (\ref{Sto2}) that the quantity governing the stationary interference is the oscillation term
\be \lb{NI}
I\propto\exp i\left[ \frac{E_2-E_1}{\hbar c}\left( x_{\!\scriptscriptstyle B}^{0}-x_{\!\scriptscriptstyle A}^{0}\right) +(\Psi_{\!\scriptscriptstyle AB})_{\scriptscriptstyle E_2, m_2} - (\Psi_{\!\scriptscriptstyle AB})_{\scriptscriptstyle E_1, m_1}\right],
\ee
with obvious notations.

For the reason invoked in Introduction, we shall henceforth suppose that $E_1=E_2$. On this assumption, the phase difference involved in Eq. (\ref{NI}) is simply the quantity $\Delta \Psi_{\!\scriptscriptstyle AB}$ defined as
\be \lb{NI2}
\Delta \Psi_{\!\scriptscriptstyle AB} = (\Psi_{\!\scriptscriptstyle AB})_{\scriptscriptstyle E, m_2} - (\Psi_{\!\scriptscriptstyle AB})_{\scriptscriptstyle E, m_1}.
\ee
This last expression is equivalent to the 
phase difference considered in Ref.~\refcite{Bhattacharya}. Using Eq. (\ref{N5b}), Eq. (\ref{NI2}) yields 
\bea \lb{n5c}
& &\Delta \Psi_{\!\scriptscriptstyle AB} = D_{\!\scriptscriptstyle AB}
\frac{(m_{2}^{2} - m_{1}^{2})c^3}{\hbar(\sqrt{E^2 - m_{1}^{2}c^4} \, + \sqrt{E^2 - m_{2}^{2}c^4})}\nonumber \\
& &\qquad\qquad\times\left[1 + \frac{E^2}{\sqrt{E^2 - m_{1}^{2}c^4} \; \sqrt{E^2 - m_{2}^{2}c^4}} {\mathcal A}({\bf x}_{\!\scriptscriptstyle A}, {\bf x}_{\!\scriptscriptstyle B})
\right].
\eea

We note that the gravitational quantity ${\mathcal B}({\bf x}_{\!\scriptscriptstyle A}, {\bf x}_{\!\scriptscriptstyle B})$ does not occurs in Eq. (\ref{n5c}). According to Eq. (\ref{B}), this feature 
shows that the gravitomagnetic components $h_{0i}$ of the metric do not contribute to the neutrino oscillations. 

We can apply the formula (\ref{n5c}) when the stationary field is treated in the slow-motion, 
post-Newtonian limit of metric theories of gravity, provided only the metric perturbations 
of order $1/c^2$ and $1/c^3$ are taken into account. Using the parametrized post-Newtonian formalism 
and assuming that the coordinates $(x^{\mu})$ correspond to a standard post-Newtonian gauge as defined in Ref.~\refcite{Will}, the relevant terms in the metric are 
\be \lb{PNds}
h_{00}=-\frac{2}{c^2}U,\qquad h_{ij}=-\frac{2\gamma}{c^2}U\delta_{ij},
\ee
where $U$ is the Newtonian gravitational potential. Then
\be \lb{15}
{\mathcal A}({\bf x}_{\!\scriptscriptstyle A}, {\bf x}_{\!\scriptscriptstyle B}) = -\frac{1}{c^2}\int_{0}^{1}U({\bf x}_{(0)}(\lambda ))d\lambda, 
\ee
where ${\bf x}_{(0)}(\lambda )$ is defined by (\ref{t9a}). Introducing the energy of the neutrinos as measured by a stationary observer at ${\bf x}_{\!\scriptscriptstyle B}$, i.e. the quantity $E_{\!\scriptscriptstyle B} =E_{loc}({\bf x}_{\!\scriptscriptstyle B})$, and then substituting for ${\mathcal A}({\bf x}_{\!\scriptscriptstyle A}, {\bf x}_{\!\scriptscriptstyle B})$ from Eq. (\ref{15}) into Eq. (\ref{n5c}) yield 
\bea 
& &\Delta \Psi_{\!\scriptscriptstyle AB} = D_{\!\scriptscriptstyle AB}\frac{(m_{2}^{2}-m_{1}^{2})c^3}{2\hbar E_{\!\scriptscriptstyle B}} \left\{ 1
+ \frac{(m_{1}^{2} + m_{2}^{2}) c^4}{4E_{\!\scriptscriptstyle B}^2}+\frac{(m_1^4+m_1^2m_2^2+m_2^4)c^8}{8E_{\!\scriptscriptstyle B}^4}+\cdots\right. \nonumber \\ 
& &\left. \qquad \quad\;\;\,+\left[1+\frac{3(m_{1}^{2} + m_{2}^{2})c^4}{4E_{\!\scriptscriptstyle B}^2}+\cdots\right]\left[\frac{U_{\!\scriptscriptstyle B}}{c^2}- \frac{1}{c^2}\int_{0}^{1}U({\bf x}_{(0)}(\lambda ))d\lambda\right] \right\} , \lb{16b}
\eea
where $D_{\!\scriptscriptstyle AB}$ is now given by 
\be \lb{DAB3}
D_{\!\scriptscriptstyle AB}=R_{\!\scriptscriptstyle AB}\left[1+\frac{\gamma}{c^2}\int_{0}^{1}U({\bf x}_{(0)}(\lambda ))d\lambda\right]
\ee
since ${\mathcal C}({\bf x}_{\!\scriptscriptstyle A}, {\bf x}_{\!\scriptscriptstyle B})=\gamma {\mathcal A}({\bf x}_{\!\scriptscriptstyle A}, {\bf x}_{\!\scriptscriptstyle B})$.

The formula (\ref{16b}) has the virtue of expressing $\Delta \Psi_{\!\scriptscriptstyle AB}$ in terms of intrinsic quantities. It will be easy to compare it with the results obtained in previous works for neutrinos propagating in a radial direction. However, the geodesic distance $D_{\!\scriptscriptstyle AB}$ has not been already introduced in the case of the propagation along a non-radial path and the constant of the motion $E$ instead of  $E_{\!\scriptscriptstyle B}$ has been often employed. So, it is useful to write Eq. (\ref{16b}) in terms of $R_{\!\scriptscriptstyle AB}$ and $E$. We find
\bea 
& &\Delta \Psi_{\!\scriptscriptstyle AB} = R_{\!\scriptscriptstyle AB}\frac{(m_{2}^{2}-m_{1}^{2})c^3}{2\hbar E} \left\{ 1
+ \frac{(m_{1}^{2} + m_{2}^{2}) c^4}{4E^2}+\frac{(m_1^4+m_1^2m_2^2+m_2^4)c^8}{8E^4}+\cdots\right. \nonumber \\ 
& &\left. \qquad \qquad +\left[\gamma -1+\frac{(\gamma -3)(m_{1}^{2} + m_{2}^{2})c^4}{4E^2}+\cdots\right]\frac{1}{c^2}\int_{0}^{1}U({\bf x}_{(0)}(\lambda ))d\lambda\right\}. \lb{16d}
\eea

Under this form, the gravitational contribution of order $(m_{2}^{2}-m_{1}^{2})c^4/E^2$ is found to be absent in general relativity, which predicts $\gamma=1$. This statement generalizes a conclusion previously drawn in Refs.~\refcite{Fornengo}, \refcite{Capozziello1}, \refcite{Crocker} and \refcite{Godunov} for radial propagations in the Schwarzschild spacetime.\footnote{Note that the gravitational contribution of order $(m_{2}^{4}-m_{1}^{4})c^8/E^4$ was calculated in Refs.~\refcite{Bhattacharya} and \refcite{Papini} under the same assumptions.} Nevertheless, we must be cautious since this conclusion is inferred from Eq. (\ref{16d})  which involves $R_{\!\scriptscriptstyle AB}$, a coordinate dependent quantity. On the other hand, it is worthy of note that the intrinsic distance $D_{\!\scriptscriptstyle AB}$ occurring in Eq. (\ref{16b}) implicitly introduces gravity even in general relativity.

Some works assume that the phase difference $\Delta\Phi_{\!\scriptscriptstyle AB}$ defined as
\be \lb{DPhi}
\Delta \Phi_{\!\scriptscriptstyle AB}=(\Phi_{\!\scriptscriptstyle AB})_{\scriptscriptstyle E, m_2} - (\Phi_{\!\scriptscriptstyle AB})_{\scriptscriptstyle E, m_1}
\ee
is the relevant quantity for determining the neutrino oscillations (see, e.g., Ref.~\refcite{Capozziello2}). So it may be of interest to mention the expression which could be deduced from Eq. (\ref{n2}) under the present assumptions. A straightforward calculation shows that 
\bea 
& &\Delta \Phi_{\!\scriptscriptstyle AB}= D_{\!\scriptscriptstyle AB}\frac{(m_{2}^{2}-m_{1}^{2})c^3}{\hbar E_{\!\scriptscriptstyle B}} \left\{ 1
+ \frac{(m_{1}^{2} + m_{2}^{2}) c^4}{2E_{\!\scriptscriptstyle B}^2}+\frac{3(m_1^4+m_1^2m_2^2+m_2^4)c^8}{8E_{\!\scriptscriptstyle B}^4}+\cdots\right. \nonumber \\ 
& &\left. \qquad \quad\;\;\,+\left[1+\frac{3(m_{1}^{2} + m_{2}^{2})c^4}{2E_{\!\scriptscriptstyle B}^2}+\cdots\right]\left[\frac{U_{\!\scriptscriptstyle B}}{c^2}- \frac{1}{c^2}\int_{0}^{1}U({\bf x}_{(0)}(\lambda ))d\lambda\right] \right\}. 
\lb{17}
\eea

We remark that $\Delta \Phi_{\!\scriptscriptstyle AB}$ is approximatively twice greater than $\Delta \Psi_{\!\scriptscriptstyle AB}$, a feature already pointed out in Ref.~\refcite{Bhattacharya}. In addition, it may be noted that the gravitational contribution of order $(m_{2}^{2}-m_{1}^{2})c^4/E^2$ in Eq. (\ref{17}) is just equal to twice the contribution of the same order appearing in Eq. (\ref{16b}).

\subsection{Neutrinos in the field of a spherically symmetric body} 

According to (\ref{16b}) and (\ref{17}), an explicit calculation of the oscillation terms $\Delta \Psi_{\!\scriptscriptstyle AB}$ or $\Delta \Phi_{\!\scriptscriptstyle AB}$ 
only requires an integration of the Newtonian potential $U$ along the line defined 
by (\ref{t9a}). As an example, we consider the gravitational field generated by an isolated, 
spherically symmetric body having a mass $M$ and a radius $r_0$. For the sake of simplicity, this body is assumed to be homogeneous. 
We shall only examine the case where the path of neutrinos is entirely outside the body and the case where 
the emission is located inside the body. We do not treat the case where the neutrinos are emitted outside 
the body but are going through the body. Indeed, the present formalism using the world function is not 
adapted to such a configuration, in which multiple-path propagations may occur. 

\subsubsection{Trajectories entirely outside the body}
In this case, $U(r)= GM/r$ at any point of the straight line defined by Eq. (\ref{t9a}). As a consequence 
\be \lb{19}
\int_{0}^{1}U({\bf x}_{(0)}(\lambda ))d\lambda =\frac{GM}{R_{\!\scriptscriptstyle AB}}\ln
\frac{r_{\!\scriptscriptstyle A}+r_{\!\scriptscriptstyle B}+R_{\!\scriptscriptstyle AB}}{r_{\!\scriptscriptstyle A}+r_{\!\scriptscriptstyle B}-R_{\!\scriptscriptstyle AB}},
\ee
where $r_{\!\scriptscriptstyle A} = \mid \! {\bf x}_{\!\scriptscriptstyle A }\! \mid $ and $r_{\!\scriptscriptstyle B} = \mid \! {\bf x}_{\!\scriptscriptstyle B} \! \mid$ (cf., e.g.,  Ref.~\refcite{Linet1} and Refs. therein). Inserting Eq. (\ref{19}) into Eq. (\ref{16b}) yields
\bea
& &\Delta \Psi_{\!\scriptscriptstyle AB} = D_{\!\scriptscriptstyle AB}\frac{(m_{2}^{2}-m_{1}^{2})c^3}{2\hbar E_{\!\scriptscriptstyle B}} \left[ 1
+ \frac{(m_{1}^{2} + m_{2}^{2}) c^4}{4E_{\!\scriptscriptstyle B}^2}+\frac{(m_1^4+m_1^2m_2^2+m_2^4)c^8}{8E_{\!\scriptscriptstyle B}^4}+\cdots\right] \nonumber \\ 
& &\qquad \qquad+\frac{GM}{c^2}\frac{(m_{2}^{2}-m_{1}^{2})c^3}{2\hbar E_{\!\scriptscriptstyle B}}\left[1+\frac{3(m_{1}^{2} + m_{2}^{2})c^4}{4E_{\!\scriptscriptstyle B}^2}+\cdots\right] \nonumber \\
& &\qquad \qquad\times\left[\frac{R_{\!\scriptscriptstyle AB}}{r_{\!\scriptscriptstyle B}}- \ln
\frac{r_{\!\scriptscriptstyle A}+r_{\!\scriptscriptstyle B}+R_{\!\scriptscriptstyle AB}}{r_{\!\scriptscriptstyle A}+r_{\!\scriptscriptstyle B}-R_{\!\scriptscriptstyle AB}}\right]. \lb{16e}
\eea

The expression for a radial propagation outside the central body is obtained by setting $R_{\!\scriptscriptstyle AB} = \vert r_{\!\scriptscriptstyle B}-r_{\!\scriptscriptstyle A}\vert$ in Eq. (\ref{16e}).

In the following subsections it will be useful to introduce the zeroth-order ``distance 
of closest approach" between the straight line passing through ${\bf x}_{\scriptscriptstyle A}$ and ${\bf x}_{\scriptscriptstyle B}$ and the origin $O$ of coordinates $x^i$, namely 
\be \lb{t20c}
r_{c} = \frac{\vert {\bf x}_{\!\scriptscriptstyle A}\times {\bf x}_{\!\scriptscriptstyle B}\vert}{R_{\!\scriptscriptstyle AB}}.
\ee

So let us express the relevant quantities in terms of $r_{\!\scriptscriptstyle A}$, $r_{\!\scriptscriptstyle B}$ and $r_{c}$. Some calculations show that
\be \lb{19b}
\frac{r_{\!\scriptscriptstyle A}+r_{\!\scriptscriptstyle B}+R_{\!\scriptscriptstyle AB}}{r_{\!\scriptscriptstyle A}+r_{\!\scriptscriptstyle B}-R_{\!\scriptscriptstyle AB}}=\frac{r_{\!\scriptscriptstyle B}+{\bf N}_{\!\scriptscriptstyle AB}\cdot{\bf x}_{\!\scriptscriptstyle B}}{r_{\!\scriptscriptstyle A}+{\bf N}_{\!\scriptscriptstyle AB}\cdot{\bf x}_{\!\scriptscriptstyle A}} ,
\ee
where ${\bf N}_{\!\scriptscriptstyle AB}=({\bf x}_{\!\scriptscriptstyle B}-{\bf x}_{\!\scriptscriptstyle A})/R_{\!\scriptscriptstyle AB}$. Then an elementary geometric reasoning shows that Eq. (\ref{19b}) reads
\be \lb{19c}
\frac{r_{\!\scriptscriptstyle A}+r_{\!\scriptscriptstyle B}+R_{\!\scriptscriptstyle AB}}{r_{\!\scriptscriptstyle A}+r_{\!\scriptscriptstyle B}-R_{\!\scriptscriptstyle AB}}=\frac{r_{\!\scriptscriptstyle B}+\epsilon'\sqrt{r_{\!\scriptscriptstyle B}^2-r_c^2}}{r_{\!\scriptscriptstyle A}+\epsilon\sqrt{r_{\!\scriptscriptstyle A}^2-r_c^2}},
\ee
where  
\be \lb{eps}
\epsilon = \mbox{sign} ({\bf x}_{\!\scriptscriptstyle A}\cdot{\bf N}_{\!\scriptscriptstyle AB}), \quad \epsilon'=\mbox{sign} ({\bf x}_{\!\scriptscriptstyle B}\cdot{\bf N}_{\!\scriptscriptstyle AB}).
\ee

Moreover, with the above definition of $\epsilon$, $R_{\!\scriptscriptstyle AB}$ is given by
\be \lb{RAB}
R_{\!\scriptscriptstyle AB}= \epsilon'\sqrt{r_{\!\scriptscriptstyle B}^2-r_c^2} - \epsilon \sqrt{r_{\!\scriptscriptstyle A}^2-r_c^2}.
\ee

Using Eqs. (\ref{19c}) and (\ref{RAB}), the last line in Eq. (\ref{16e}) may be rewritten as
\be \nonumber
\frac{R_{\!\scriptscriptstyle AB}}{r_{\!\scriptscriptstyle B}}- \ln
\frac{r_{\!\scriptscriptstyle A}+r_{\!\scriptscriptstyle B}+R_{\!\scriptscriptstyle AB}}{r_{\!\scriptscriptstyle A}+r_{\!\scriptscriptstyle B}-R_{\!\scriptscriptstyle AB}}= \frac{\epsilon'\sqrt{r_{\!\scriptscriptstyle B}^2-r_c^2} - \epsilon \sqrt{r_{\!\scriptscriptstyle A}^2-r_c^2}}{r_{\!\scriptscriptstyle B}}- \ln\frac{r_{\!\scriptscriptstyle B}+\epsilon'\sqrt{r_{\!\scriptscriptstyle B}^2-r_c^2}}{r_{\!\scriptscriptstyle A}+\epsilon\sqrt{r_{\!\scriptscriptstyle A}^2-r_c^2}}. 
\ee

It may be noted that the expression for a radial propagation is now obtained by setting $r_c = 0$.

\subsubsection{Emission located inside the massive body and reception located outside}
In order to simplify the evaluation of the integral in the R.H.S. of (\ref{15}), it is useful to introduce the new parameter $\overline{\lambda}$ defined by 
\be \nonumber
\overline{\lambda} = \lambda + \frac{{\bf N}_{\!\scriptscriptstyle AB}\cdot{\bf x}_{\!\scriptscriptstyle A}}{R_{\!\scriptscriptstyle AB}},
\ee
the range of $\overline{\lambda}$ being $\overline{\lambda}_{\!\scriptscriptstyle A} \leq \overline{\lambda} 
\leq \overline{\lambda}_{\!\scriptscriptstyle B}$, with $\overline{\lambda}_{\!\scriptscriptstyle A} = {\bf N}_{\!\scriptscriptstyle AB}\cdot{\bf x}_{\!\scriptscriptstyle A}/{R_{\!\scriptscriptstyle AB}}$ and $\overline{\lambda}_{\!\scriptscriptstyle B} = {\bf N}_{\!\scriptscriptstyle AB}\cdot{\bf x}_{\!\scriptscriptstyle B}/{R_{\!\scriptscriptstyle AB}}$. Denoting by $\overline{\lambda}_{0}$ the value of $\overline{\lambda}$ corresponding to the intersection 
of the segment connecting ${\bf x}_{\!\scriptscriptstyle A}$ and ${\bf x}_{\!\scriptscriptstyle B}$ with the sphere $r= r_{0}$, we have  
\be \lb{intU2}
\int_{0}^{1} U({\bf x}_{(0)}(\lambda)) d\lambda = 
\int_{\overline{\lambda}_{\!\scriptscriptstyle A}}^{\overline{\lambda}_{0}} \frac{GM}{2 r_0} \left[3 - \frac{\vert {\bf x}_{(0)}(\overline{\lambda})\vert^2}{r_0^2} \right]  
d\overline{\lambda} + 
\int_{\overline{\lambda}_{0}}^{\overline{\lambda}_{\!\scriptscriptstyle B}} \frac{GM}{\mid \! 
{\bf x}_{(0)}(\overline{\lambda}) \! \mid} 
d\overline{\lambda}. 
\ee
Using $\overline{\lambda}_{0} = \sqrt{r_0^2 - r_c^2}/R_{\!\scriptscriptstyle AB}$, a lengthy but straightforward calculation leads to  
\bea \lb{21}
& &\Delta \Psi_{\!\scriptscriptstyle AB} = D_{\!\scriptscriptstyle AB}\frac{(m_{2}^{2}-m_{1}^{2})c^3}{2\hbar E_{\!\scriptscriptstyle B}} \left[ 1
+ \frac{(m_{1}^{2} + m_{2}^{2}) c^4}{4E_{\!\scriptscriptstyle B}^2}+\frac{(m_1^4+m_1^2m_2^2+m_2^4)c^8}{8E_{\!\scriptscriptstyle B}^4}+\cdots\right] \nonumber \\ 
& &\qquad \qquad +\frac{GM}{c^2}\frac{(m_{2}^{2}-m_{1}^{2})c^3}{2\hbar E_{\!\scriptscriptstyle B}}\left[1+\frac{3(m_{1}^{2} + m_{2}^{2})c^4}{4E_{\!\scriptscriptstyle B}^2}+\cdots\right] \nonumber \\
& &\qquad\qquad \times \Bigg\lbrack \frac{R_{\!\scriptscriptstyle AB}}{r_{\!\scriptscriptstyle B}} -\frac{\sqrt{r_{0}^{2}-r_{c}^{2}} - \epsilon \sqrt{r_{\!\scriptscriptstyle A}^{2}-r_{c}^{2}}}{r_0}
\left( \frac{4}{3}-\frac{r_{\!\scriptscriptstyle A}^{2}+r_{c}^{2} + \epsilon \sqrt{r_{0}^{2}-r_{c}^{2}}
\sqrt{r_{\!\scriptscriptstyle A}^{2}-r_{c}^{2}}}{6r_{0}^{2}}\right)  \nonumber \\
& &\qquad \qquad - \ln 
\left( \frac{r_{\!\scriptscriptstyle B}+\sqrt{r_{\!\scriptscriptstyle B}^2-r_c^2}}{r_0+\sqrt{r_{0}^{2}-r_{c}^{2}}}\right)\Bigg\rbrack,
\eea
where $R_{\!\scriptscriptstyle AB}$ may be replaced by the R.H.S. of Eq. (\ref{RAB}) with $\epsilon'=1$.

\subsection{Comparison with previous results for non-radial propagation}

Among all the above-mentioned papers, the most coherent study of the gravitational quantum phase shift for a non-radial timelike geodesic in an exterior Schwarzschild spacetime is given in Ref.~\refcite{Crocker}.\footnote{Some authors carry out their calculations along null geodesics (see, e.g., Refs.~\refcite{Fornengo} and \refcite{Zhang}).} In what follows, we want to compare the expression of $\Psi_{\!\scriptscriptstyle AB}$ found in this work for a non-radial trajectory with our  formula (\ref{N5}) written outside a spherically symmetric massive body. To do this, we have to calculate $R_{\!\scriptscriptstyle AB}$ as a function of $r_{\!\scriptscriptstyle A}$, $r_{\!\scriptscriptstyle B}$ and the true value $r_{\!\scriptscriptstyle P}$ of the radial coordinate at the pericenter of the trajectory. We can proceed to this calculation by determining the relation between $r_{\!\scriptscriptstyle P}$ and $r_c$. It will be assumed that $r_{\!\scriptscriptstyle P} \gg GM/c^2$.

In isotropic spherical coordinates $(x^0, r, \vartheta, \varphi)$, the metric (\ref{PNds}) may be written in the form
\be \lb{PNds2}
ds^2 = \left(1 - \frac{2GM}{c^2r}\right)(dx^0)^2 - \left(1 + \frac{2\gamma GM}{c^2r}\right)(dr^2 + r^2 d\vartheta^2 + r^2 \sin^2 \vartheta d\varphi^2).
\ee
It is well known that the trajectory is in a plane passing through the origin $O$. As a consequence, the axes may be chosen  in such a way that  $\vartheta = \pi/2$ during the propagation. Then, using the Euler-Lagrange equations for the geodesics of (\ref{PNds2}), it is easily seen that the path of a particle of mass $m$ is determined by an equation as follows
\be \lb{traj}
\left(\frac{dr}{d\varphi}\right)^2 = r^2\left[\frac{r^2}{b^2} + \frac{2GM}{c^2 b}\left(\gamma+\frac{E^2}{E^2 - m^2 c^4}\right)\frac{r}{b} - 1\right],
\ee
where $b$ is the impact parameter of the particle, defined as the Euclidean distance which could be measured by an inertial observer at rest at infinity between the asymptote to the ray and the line parallel to this asymptote passing through the center $O$. The pericenter $P$ being defined as the point where $dr/d\varphi = 0$, the value of $r$ at this point is immediately deduced from Eq. (\ref{traj}). It is easily seen that $b$ and $r_{\!\scriptscriptstyle P}$ are related by
\be \lb{brP}
b = r_{\!\scriptscriptstyle P} \sqrt{1+\left(\gamma+\frac{E^2}{E^2 - m^2 c^4}\right)\frac{2GM}{c^2r_{\!\scriptscriptstyle P}}}.
\ee
Since we suppose that $r_{\!\scriptscriptstyle P} \gg GM/c^2$, Eq. (\ref{brP}) reduces to 
\be \lb{rP}
b =r_{\!\scriptscriptstyle P} + \left(\gamma+\frac{E^2}{E^2 - m^2 c^4}\right)\frac{GM}{c^2}.
\ee

Let us now proceed to the calculation of the impact parameter $b$ as a function of $r_{\!\scriptscriptstyle A}$, $r_{\!\scriptscriptstyle B}$ and $r_c$. Returning to the quasi Cartesian isotropic coordinates used in Eqs. (\ref{PNds}), and introducing the vector notation
$\underline{{\bf p}} = (p_1, p_2, p_3)$, where the quantities $p_i$ are the spacelike covariant components of the energy-momentum 4-vector defined in Sect. II, it may be seen from the geodesic equations that the vector ${\bf L}$ defined as
${\bf L} = - {\bf x} \times \underline{{\bf p}}$
remains constant along any geodesic path. For a freely falling particle able to go to infinity, the magnitude of ${\bf L}$ is such that
$\vert {\bf L}\vert = \lim_{\vert {\bf x}\vert \rightarrow \infty}\left\vert - {\bf x}\times \underline{{\bf p}}\right\vert =b E/c^2 \vert (d{\bf x}/dt)_{\infty}\vert$ since $\underline{{\bf p}} \longrightarrow -(E/c^2)(d{\bf x}/dt)_{\infty}$ when $\vert{\bf x}\vert \longrightarrow \infty$. But it follows from Eq. (\ref{Ev}) that $\vert (d{\bf x}/cdt)_{\infty}\vert = \sqrt{1 - m^2c^4/E^2}$. As a consequence the impact parameter $b$ may be identified to the constant of the motion given by\footnote{This equation generalizes the relation beween the conserved angular momentum ${\bf L}$ and the impact parameter for a photon (see, e.g., Ref.~\refcite{Chandrasekhar}).}
\begin{equation} \label{imp2}
b = \frac{c}{\sqrt{E^2 - m^2c^4}}\vert\!-{\bf x}\times \underline{{\bf p}}\,\vert .
\end{equation}

According to Eqs. (\ref{2a}), (\ref{t2c}) and (\ref{Sto2}), we have $(\underline{{\bf p}})_{\!\scriptscriptstyle B} = \hbar {\bm \nabla}_{{\bf x}_{\!\scriptscriptstyle B}}\Psi_{\!\scriptscriptstyle AB}$. As a consequence, taking Eq. (\ref{N5}) into account, we get
\bea 
& &(\underline{{\bf p}})_{\!\scriptscriptstyle B} = - \frac{1}{c}\sqrt{E^2 - m^2c^4} \nonumber \\
& &\qquad\quad\,\times\bigg\lbrack {\bf N}_{\!\scriptscriptstyle AB} + \left(\gamma+\frac{E^2}{E^2 - m^2 c^4}\right)\frac{GM}{c^2} \frac{(r_{\!\scriptscriptstyle A} + r_{\!\scriptscriptstyle B}) {\bf N}_{\!\scriptscriptstyle AB} - R_{\!\scriptscriptstyle AB} {\bf n}_{\!\scriptscriptstyle B}}{r_{\!\scriptscriptstyle A} r_{\!\scriptscriptstyle B}(1 + {\bf n}_{\!\scriptscriptstyle A}\cdot{\bf n}_{\!\scriptscriptstyle B})}\bigg\rbrack,\lb{imp3}
\eea
where ${\bf n}_{\!\scriptscriptstyle A} = {\bf x}_{\!\scriptscriptstyle A}/r_{\!\scriptscriptstyle A}$ and ${\bf n}_{\!\scriptscriptstyle B} =  {\bf x}_{\!\scriptscriptstyle B}/r_{\!\scriptscriptstyle B}$. \lb{p3}

Using Eq. (\ref{imp3}), and then taking into account Eq. (\ref{t20c}), Eq. (\ref{imp2}) gives an expression as follows for the impact parameter
\be \lb{imp4}
b = r_c \left[1 + \left(\gamma+\frac{E^2}{E^2 - m^2 c^4}\right)\frac{GM}{c^2}
\frac{r_{\!\scriptscriptstyle A} + r_{\!\scriptscriptstyle B}}{r_{\!\scriptscriptstyle A} r_{\!\scriptscriptstyle B}(1 + {\bf n}_{\scriptscriptstyle A}\cdot{\bf n}_{\scriptscriptstyle B})}\right].
\ee
Comparing now Eq. (\ref{imp4}) with Eq. (\ref{rP}) shows that $r_{\!\scriptscriptstyle P}$ and $r_{c}$ are linked by the relation
\be \lb{rP2}
r_{\!\scriptscriptstyle P} = r_c + \left(\gamma+\frac{E^2}{E^2 - m^2 c^4}\right)\frac{GM}{c^2 r_c}\left[\frac{r_c^2 (r_{\!\scriptscriptstyle A} + r_{\!\scriptscriptstyle B})}{r_{\!\scriptscriptstyle A} r_{\!\scriptscriptstyle B}(1 + {\bf n}_{\scriptscriptstyle A}\cdot{\bf n}_{\scriptscriptstyle B})} - r_c\right].
\ee

For shortness, we limit ourselves to the case $r_{\!\scriptscriptstyle B} > r_{\!\scriptscriptstyle A}$, which implies $\epsilon'=1$. Noting that Eq. (\ref{t20c}) implies $r_c^2 = r_{\!\scriptscriptstyle A}^2 r_{\!\scriptscriptstyle B}^2 [1 - ({\bf n}_{\scriptscriptstyle A}\cdot{\bf n}_{\scriptscriptstyle B})^2]/R_{\!\scriptscriptstyle AB}^2$ and that
\bea 
& &r_{\!\scriptscriptstyle A} - r_{\!\scriptscriptstyle B} ({\bf n}_{\scriptscriptstyle A}\cdot{\bf n}_{\scriptscriptstyle B}) = - \epsilon 
\frac{R_{\!\scriptscriptstyle AB}}{r_{\!\scriptscriptstyle A}}\sqrt{r_{\!\scriptscriptstyle A}^2 - r_c^2}, \nonumber \\
& &r_{\!\scriptscriptstyle B} - r_{\!\scriptscriptstyle A} ({\bf n}_{\scriptscriptstyle A}\cdot{\bf n}_{\scriptscriptstyle B}) = \frac{R_{\!\scriptscriptstyle AB}}{r_{\!\scriptscriptstyle B}}\sqrt{r_{\!\scriptscriptstyle B}^2 - r_c^2}, \nonumber
\eea
it may be seen that Eq. (\ref{rP2}) becomes
\be \lb{rP3}
r_{\!\scriptscriptstyle P} = r_c + \left(\gamma+\frac{E^2}{E^2 - m^2 c^4}\right)\frac{GM}{c^2 r_c} \frac{(r_{\!\scriptscriptstyle A} - r_c)\sqrt{r_{\!\scriptscriptstyle B}^2 - r_c^2} - \epsilon (r_{\!\scriptscriptstyle B} - r_c)\sqrt{r_{\!\scriptscriptstyle A}^2 - r_c^2}}{\sqrt{r_{\!\scriptscriptstyle B}^2 - r_c^2}-\epsilon \sqrt{r_{\!\scriptscriptstyle A}^2 - r_c^2}}.
\ee

Now, inverting Eq. (\ref{rP3}) in order to find $r_c$ as a function of $r_{\!\scriptscriptstyle A}$, $r_{\!\scriptscriptstyle B}$ and $r_{\!\scriptscriptstyle P}$, and then substituting the result into Eq. (\ref{RAB}) yield 
\bea \lb{RAB2}
& &R_{\!\scriptscriptstyle AB} =  \sqrt{r_{\!\scriptscriptstyle B}^2-r_{\!\scriptscriptstyle P}^2} - \epsilon \sqrt{r_{\!\scriptscriptstyle A}^2-r_{\!\scriptscriptstyle P}^2} \nonumber \\
& &\qquad\quad+  \left(\gamma+\frac{E^2}{E^2 - m^2 c^4}\right)\frac{GM}{c^2} \left(\sqrt{\frac{r_{\!\scriptscriptstyle B} - r_{\!\scriptscriptstyle P}}{r_{\!\scriptscriptstyle B} + r_{\!\scriptscriptstyle P}}} - \epsilon \sqrt{\frac{r_{\!\scriptscriptstyle A} - r_{\!\scriptscriptstyle P}}{r_{\!\scriptscriptstyle A} + r_{\!\scriptscriptstyle P}}}\right).
\eea

Substituting for $R_{\!\scriptscriptstyle AB}$ from Eq. (\ref{RAB2}) into Eq. (\ref{N5}) gives for $\Psi_{\!\scriptscriptstyle AB}$:
\bea 
& &\Psi_{\!\scriptscriptstyle AB}=-\frac{\sqrt{E^2-m^2c^4}}{\hbar c} \Bigg\lbrace\sqrt{r_{\!\scriptscriptstyle B}^2-r_{\!\scriptscriptstyle P}^2} - \epsilon \sqrt{r_{\!\scriptscriptstyle A}^2-r_{\!\scriptscriptstyle P}^2} + \left(\gamma+\frac{E^2}{E^2 - m^2 c^4}\right)\frac{GM}{c^2} \nonumber \\
& &\qquad\quad\times\left[\sqrt{\frac{r_{\!\scriptscriptstyle B} - r_{\!\scriptscriptstyle P}}{r_{\!\scriptscriptstyle B} + r_{\!\scriptscriptstyle P}}} - \epsilon \sqrt{\frac{r_{\!\scriptscriptstyle A} - r_{\!\scriptscriptstyle P}}{r_{\!\scriptscriptstyle A} + r_{\!\scriptscriptstyle P}}} + \ln\frac{r_{\!\scriptscriptstyle B} + \sqrt{r_{\!\scriptscriptstyle B}^2-r_{\!\scriptscriptstyle P}^2}}{r_{\!\scriptscriptstyle A} + \epsilon \sqrt{r_{\!\scriptscriptstyle A}^2-r_{\!\scriptscriptstyle P}^2}} \right]\Bigg\rbrace. \lb{N6}
\eea

The change of radial coordinates defined as 
\be \nonumber 
r=\rho - \frac{\gamma GM}{c^2} 
\ee  
transforms the metric (\ref{PNds2}) into
\be \lb{Schw2}
ds^2 = \left( 1 - \frac{2GM}{c^2 \rho} \right) - \left(1 + \frac{2 \gamma GM}{c^2 \rho}\right)d\rho^2 - \rho^2(d\theta^2 + \sin^2\theta d\phi^2)
\ee
up to terms of order $G^2$. This metric coincides of course with the linearized Schwarzschild metric when $\gamma = 1$.

Noting that 
\be \nonumber
\sqrt{r^2 - r_{\!\scriptscriptstyle P}^2} = \sqrt{\rho^2 - \rho_{\!\scriptscriptstyle P}^2} - \gamma \frac{GM}{c^2} \sqrt{\frac{\rho - \rho_{\!\scriptscriptstyle P}}{\rho + \rho_{\!\scriptscriptstyle P}}}
\ee
for any $r \geq r_{\!\scriptscriptstyle P}$, and then neglecting terms of order $m^4c^8/E^4$ Eq. (\ref{N6})  finally gives
\bea 
& &\Psi_{\!\scriptscriptstyle AB} = \Psi^{\ast}_{\!\scriptscriptstyle AB}
+\frac{m^2c^3}{2\hbar E}\Bigg\lbrace \sqrt{\rho_{\!\scriptscriptstyle B}^2 - \rho_{\!\scriptscriptstyle P}^2} - \epsilon \sqrt{\rho_{\!\scriptscriptstyle A}^2 - \rho_{\!\scriptscriptstyle P}^2} \nonumber \\
& &\qquad\quad- \frac{GM}{c^2}\Bigg\lbrack
\sqrt{\frac{\rho_{\!\scriptscriptstyle B} - \rho_{\!\scriptscriptstyle P}}{\rho_{\!\scriptscriptstyle B} + \rho_{\!\scriptscriptstyle P}}} - \epsilon \sqrt{\frac{\rho_{\!\scriptscriptstyle A} - \rho_{\!\scriptscriptstyle P}}{\rho_{\!\scriptscriptstyle A} + \rho_{\!\scriptscriptstyle P}}} - (\gamma - 1)
\ln\frac{\rho_{\!\scriptscriptstyle B} + \sqrt{\rho_{\!\scriptscriptstyle B}^2-\rho_{\!\scriptscriptstyle P}^2}}{\rho_{\!\scriptscriptstyle A} + \epsilon \sqrt{\rho_{\!\scriptscriptstyle A}^2-\rho_{\!\scriptscriptstyle P}^2}}\Bigg\rbrack\Bigg\rbrace, \qquad \lb{rrho}
\eea
where $\Psi_{\!\scriptscriptstyle AB}^{\star} $ is a quantity which does not depend on the mass $m$ and consequently disappears from the expression of $\Delta\Psi_{\!\scriptscriptstyle AB}$:
\bea 
& &\Psi_{\!\scriptscriptstyle AB}^{\star}  = -\frac{E}{\hbar c} \Bigg\lbrace \sqrt{\rho_{\!\scriptscriptstyle B}^2 - \rho_{\!\scriptscriptstyle P}^2} - \epsilon \sqrt{\rho_{\!\scriptscriptstyle A}^2 - \rho_{\!\scriptscriptstyle P}^2} + \frac{GM}{c^2}\Bigg\lbrack
\sqrt{\frac{\rho_{\!\scriptscriptstyle B} - \rho_{\!\scriptscriptstyle P}}{\rho_{\!\scriptscriptstyle B} + \rho_{\!\scriptscriptstyle P}}} - \epsilon \sqrt{\frac{\rho_{\!\scriptscriptstyle A} - \rho_{\!\scriptscriptstyle P}}{\rho_{\!\scriptscriptstyle A} + \rho_{\!\scriptscriptstyle P}}}\nonumber \\
& &\qquad\quad\,+ (\gamma + 1)
\ln\frac{\rho_{\!\scriptscriptstyle B} + \sqrt{\rho_{\!\scriptscriptstyle B}^2-\rho_{\!\scriptscriptstyle P}^2}}{\rho_{\!\scriptscriptstyle A} + \epsilon \sqrt{\rho_{\!\scriptscriptstyle A}^2-\rho_{\!\scriptscriptstyle P}^2}}\Bigg\rbrack\Bigg\rbrace. \lb{rrh2}
\eea

Putting $\gamma = 1$, we recover Eqs. (51) and (52) in Ref.~\refcite{Crocker} according as $\epsilon = 1$ or $\epsilon = -1$.

\section{Conclusion} 

Synge's world function constitutes a powerful tool for determining the quantum phase 
shift of freely falling particles in any stationary, weak gravitational field considered in the linear approximation. The method 
works provided the spin effects may be neglected and multiple-path effects are not relevant.

It is shown by the formulas (\ref{16b}) and (\ref{16d}) that the neutrino oscillations are not sensitive to the gravitomagnetic components of the metric and that the post-Newtonian parameter $\gamma$ is involved in the gravitational phase shift only via the expression 
of the geodesic distance $D_{\!\scriptscriptstyle AB}$. 

Explicit formulas are written for neutrinos propagating in the field of a homogeneous, spherically symmetric body, see Eqs. (\ref{16e}) and (\ref{21}). The calculations are straightforward even for neutrinos emitted inside 
the body and traveling along non-radial geodesics. It is shown that the results previously obtained in Schwarzschild spacetime are direct consequences of our more general formulas. Moreover, it must be pointed that our results are given in terms of the position vectors ${\bf x}_{\!\scriptscriptstyle A}$ and ${\bf x}_{\!\scriptscriptstyle B}$, in contrast with the formulas obtained in Ref.~\refcite{Crocker}, where $r_{\!\scriptscriptstyle P}$ is not explicitly calculated.

\section*{Acknowledgments}

We are grateful to Prof. A. D. Ahluwalia for having encouraged us to publish these results.


\begin{thebibliography}{0}
\bibitem{Ahluwalia1} D. V. Ahluwalia and C. Burgard, {\it Gen. Rel. Grav.} {\bf 28}, 1161 
(1996).
\bibitem{Ahluwalia2} D. V. Ahluwalia and C. Burgard, {\it Phys. Rev. D} {\bf 57}, 4724 (1998).
\bibitem{Fornengo} N. Fornengo, C. Giunti, C. W. Kim, and J. Song, {\it Phys. Rev. D} {\bf 56}, 1895 (1997).
\bibitem{Bhattacharya} T. Bhattacharya, S. Habib, and E. Mottola, {\it Phys. Rev. D} {\bf 59}, 067301 (1999).
\bibitem{Capozziello1} S. Capozziello and G. Lambiase, {\it Mod. Phys. Lett. A} {\bf 14}, 2193 (1999).
\bibitem{Wudka} J. Wudka, {\it Phys. Rev. D} {\bf 64}, 065009 (2001).
\bibitem{Crocker} R. M. Crocker, C. Giunti and D. J. Mortlock, {\it Phys. Rev. D} {\bf 69}, 063008 (2004).
\bibitem{Godunov} S. I. Godunov and G. S. Pastukhov, {\it Phys. Atom. Nucl.} {\bf 74}, 302 (2011).
\bibitem{Capozziello2} S. Capozziello, M. De Laurentis and D. Vernieri, {\it Mod. Phys. Lett. A} {\bf 25}, 1163 (2010).
\bibitem{Synge} J. L. Synge, {\it Relativity: The General Theory}
(North-Holland, 1964).
\bibitem{Linet1} B. Linet and P. Teyssandier, {\it Phys. Rev. D} {\bf 66}, 024045 (2002).
\bibitem{Linet2} C. Le Poncin-Lafitte, B. Linet and P. Teyssandier, {\it Class. Quantum Grav.} {\bf 21}, 4463 (2004).
\bibitem{Teyssandier} P. Teyssandier and C. Le Poncin-Lafitte, {\it Class. Quantum Grav.} {\bf 25}, 145020 (2008).
\bibitem{Piriz} D. P\'{\i}riz, M. Roy, and J. Wudka, {\it Phys. Rev. D} {\bf 54}, 1587 (1996).
\bibitem{Cardall} C. Y. Cardall and G. M. Fuller, {\it Phys. Rev. D} {\bf 55}, 7960 (1997).
\bibitem{Stodolsky1} L. Stodolsky, {\it Phys. Rev. D} {\bf 58}, 036006 (1998).
\bibitem{Lipkin} H. J. Lipkin, {\it Phys. Lett. B} {\bf 642}, 366 (2006).
\bibitem{Stodolsky2} L. Stodolsky, {\it Gen. Rel. Grav.} {\bf 11}, 391 (1979). 
\bibitem{Nandi} K. K. Nandi and Yuan-Zhong Zhang, {\it Phys. Rev. D} {\bf 66}, 063005 (2002).
\bibitem{Landau} L. Landau and E. Lifchitz, {\it The Classical Theory of Fields} (Butterworth-Heinemann, 4th rev. edition, 1987).
\bibitem{Will} C. M. Will, {\it Theory and Experiment in Gravitational Physics} (Cambridge University Press, 1981).
\bibitem{Papini} G. Papini, in {\it  Advances in the Interplay between
Quantum and Gravity Physics, Proc. of the NATO Advanced Study Institute} (Erice, 2001), eds.~P. G. Bergmann and V. Sabbata (Kluwer, 2002), p. 317; arXiv:gr-qc/0110056.
\bibitem{Zhang} C. M. Zhang and A. Beesham, {\it Int. J. Mod. Phys. D} {\bf 12}, 727 (2003).
\bibitem{Chandrasekhar} S. Chandrasekhar, {\it The Mathematical Theory of Black Holes} (Clarendon Press, 1983).






\end{thebibliography}
\end{document}